\documentclass[useAMS,usenatbib]{mn2e}

\usepackage{graphicx}

\title[Cross-entropy optimiser]{Cross-entropy optimiser: a new tool to study precession in astrophysical jets}
\author[A. Caproni, H. Monteiro and Z. Abraham]{A. Caproni$^{1}$\thanks{E-mail:
anderson.caproni@cruzeirodosul.edu.br}, H. Monteiro$^{1}$ and Z. Abraham$^{2}$\\
$^{1}$N\'ucleo de Astrof\'\i sica Te\'orica, Universidade Cruzeiro do Sul, R. Galv\~ao Bueno 868, Liberdade, 01506-000, S\~ao Paulo, SP, Brazil\\
CEP 05508-900, S\~ao Paulo, SP, Brazil\\
$^{2}$Instituto de Astronomia, Geof\'\i sica e Ci\^encias Atmosf\'ericas, Universidade de S\~ao Paulo, R. do Mat\~ao 1226, Cidade Universit\'aria,\\ 
CEP 05508-900, S\~ao Paulo, SP, Brazil}

\begin{document}

\date{Submitted}

\pagerange{\pageref{firstpage}--\pageref{lastpage}} \pubyear{2002}

\maketitle

\label{firstpage}

\begin{abstract}
Evidence of jet precession in many galactic and extragalactic sources has been reported in the literature. Much of this evidence is based on studies of the kinematics of the jet knots, which depends on the correct  identification of the components to determine their respective proper motions and position angles on the plane of the sky. Identification problems related to fitting procedures, as well as observations poorly sampled in time, may influence the follow up of the components in time, which consequently might contribute to a misinterpretation of the data. In order to deal with these limitations, we introduce a very powerful statistical tool to analyse jet precession: the cross-entropy method for continuous multi-extremal optimisation. Only based on the raw data of the jet components (right ascension and declination offsets from the core), the cross-entropy method searches for the precession model parameters that better represent the data. In this work we present a large number of tests to validate this technique, using synthetic precessing jets built from a given set of precession parameters. Aiming to recover these parameters, we applied  the cross-entropy method to our precession model, varying exhaustively the quantities associated to the method. Our results have shown that even in the most challenging tests, the cross-entropy method was able to find the correct parameters within $1\%$-level. Even for a non-precessing jet, our optimization method could point out successfully the lack of precession. 
\end{abstract}


\section{Introduction}

Improvements in the sensitivity and angular resolution of astronomical instruments have allowed the study of galactic and extragalactic jets in relative detail. In our Galaxy, jets have been observed directly in star-forming regions (e.g., \citealt{rei04}), planetary nebula (e.g., \citealt{hugg07}) and in micro-quasars (e.g., \citealt{sti02}). The Galactic Centre might also harbour a very faint jet generated in the vicinity of a super-massive black hole (e.g., \citealt{kri93}). In the case of extragalactic sources, jets have been detected in the cores of active galaxies, some of them having relativistic speeds (e.g., \citealt{zens97}). Discrete components, which nature is still a matter of debate, are often seen receding from the unresolved core, where the engine responsible for their acceleration is supposed to be located (e.g., \citealt{jun99}).

The accumulation of observational data  in a wide range of radio frequencies have shown that a large number of jets are not perfectly linear, exhibiting a bent shape that has been  interpreted as due to jet-ambient interaction (e.g., \citealt{gal04}), hydro- or magnetohydrodynamical instabilities (e.g., \citealt{ferr98}), jet rotation (e.g., \citealt{cego04}), jet inlet precession (e.g., \citealt{caab04a,caab04b}) or a combination of them (e.g., \citealt{har01}). 

Concerning jet precession, \citet{abca98} developed an analytical model to study precession in the quasar 3C\,279, which was also applied to the quasars 3C\,273 and 3C\,345 \citep{abro99,caab04a},  the BL Lac object OJ\,287 and the Seyfert galaxy 3C\,120 \citep{abra00,caab04b}. The precession model parameters in those works were estimated from the position angles, velocities, and epoch of formation of the superluminal features in the radio jet, as well as from the long-term periodic variability at optical wavelengths in the case of OJ\,287, 3C\,345 and 3C\,120.

In order to study jet precession, it is necessary to monitor the motion of jet knots, collecting data during a long enough period of time (two or more precession periods), with the interval between consecutive observations as short as possible. The reason for this is to avoid misidentification of the jet components, which can lead to the wrong determination of their kinematic parameters (e.g., proper motion), as well as to the wrong estimation of the precession parameters. However, note that such ideal monitoring is hardly achieved in practical situations.

Aiming to overcome, or at least minimise the referred problems related to the jet precession analyses, we introduce a powerful statistical technique recently developed to deal with multi-extremal problems involving optimization: the cross-entropy method (hereafter CE).

The CE analysis was originally used in the optimization of complex computer simulation models involving rare events simulations \citep{rubi97}, having been modified by \citet{rubi99} to deal with continuous multi-extremal and discrete combinatorial optimization problems. Its theoretical asymptotic convergence has been demonstrated by \citet{marg04}, while \citet{kro06} studied the efficiency of the CE method in solving continuous multi-extremal optimization problems. Some examples of robustness of the CE method in several situations are listed in \citet{deb05}. 

The basic procedures involved in the CE optimization can be summarised as follows (e.g., \citealt{kro06}):

\begin{enumerate}
  \item Random generation of the initial parameter sample, obeying predefined criteria;
  \item Selection of the best samples based on some mathematical criterion;
  \item Random generation of updated parameter samples from the previous best candidates to be evaluated in the next iteration;
  \item Optimization process repeats steps (ii) and (iii) until a pre-specified stopping criterion is fulfilled.   
\end{enumerate}

In this work we validate our CE jet precession model from a variety of benchmark tests built from synthetic precessing jet components. They show the great capability of our method to determine jet kinematic parameters without any identification scheme for the jet knots, as usually made in the literature. This paper is structured as follows: in $\S$ 2, we introduce the CE algorithm and the jet precession model used in the calculations. Validation tests and the CE parameters are also presented in this section. The optimization results for each benchmark test of $\S$ 2, as well as for three complementary tests are discussed in $\S$ 3. Conclusions are presented in $\S$ 4.

\section{The cross-entropy method and the jet precession}

In this section we introduce the cross-entropy method, describing how to estimate the precession model parameters from the right ascension and 
declination offsets between the jet components and the unresolved core obtained from the model fittings of radio interferometric observations.

   \begin{figure}
	  {\includegraphics[width = 60 mm, height = 60 mm]{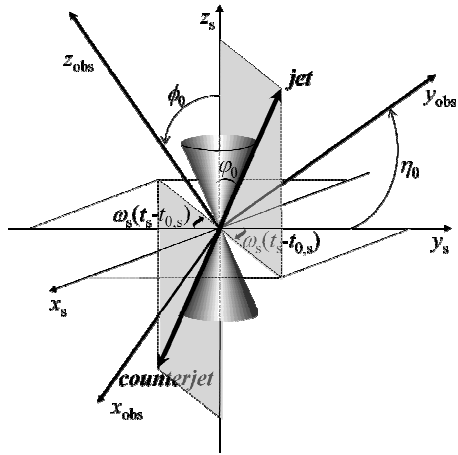}}
      \caption{Jet and counter jet precession in the source's reference frame (see text for the meaning of the symbols). The axis $z_\rmn{obs}$ points to the observer.}
      \label{precession model configuration}
   \end{figure}

\subsection{Cross-entropy algorithm for continuous optimization}

Let us suppose that we wish to study a set of $N_\rmn{d}$ observational data in terms of an analytical model characterized by $N_\rmn{p}$ parameters $p_1, p_2, ..., p_{N_\rmn{p}}$.

The main goal of the CE continuous multi-extremal optimization method is to find the set of parameters ${\bf x}^*=(p^*_1,p^*_2,...,p^*_{Np})$ for which the model provides the best description of the data \citep{rubi99,kro06}. It is performed generating randomly $N$ independent sets of model parameters ${\bf X}=({\bf x}_1,{\bf x}_2,...,{\bf x}_N)$, where ${\bf x}_i=(p_{1i},p_{2i},...,p_{N{\rmn{p}i}})$, and minimizing an objective function $S({\bf x})$ used to transmit the quality of the fit during the run process. If the convergence to the exact solution is achieved then $S({\bf x}^*)\rightarrow 0$.

In order to find the optimal solution from CE optimization, we start by defining the parameter range in which the algorithm will search for the best candidates: $p^\rmn{min}_j\leq p_j(k) \leq p^\rmn{max}_j$, where $k$ represents the iteration number. Introducing $\bar{p}_j(0)=(p^\rmn{min}_j+p^\rmn{max}_j)/2$ and $\sigma_j(0)=(p^\rmn{max}_j-p^\rmn{min}_j)/2$, we can compute ${\bf X}(0)$ from:

\begin{equation}
  X_{ij}(0)=\bar{p}_j(0)+\sigma_j(0) G_{ij},
\end{equation}
where $G_{ij}$ is an $N\times N_\rmn{p}$ matrix with random numbers generated from a zero-mean normal distribution with standard deviation of unity.

The next step is to calculate $S_i(0)$ for each set of ${\bf x}_i(0)$, ordering them according to increasing values of $S_i$. Then the  first $N_\rmn{elite}$\footnote{The estimation of the optimal or near-optimal value for the parameter $N_\rmn{elite}$ for our validation tests is discussed in Section 3.} set of parameters is selected, i.e. the $N_\rmn{elite}$-samples with lowest $S$-values, which will be labeled as the elite sample array ${\bf X}^\rmn{elite}(0)$.

We then determine  the mean and standard deviation of the elite sample, $\bar{p}^\rmn{elite}_j(0)$ and ${\bf \sigma}^\rmn{elite}_j(0)$ respectively, as:

\begin{equation}
  \bar{p}^\rmn{elite}_j(0)=\frac{1}{N_\rmn{elite}}\sum\limits_{i=1}^{N_\rmn{elite}}X^\rmn{elite}_{ij}(0),
\end{equation}

\begin{equation}
  {\bf \sigma}^\rmn{elite}_j(0)=\sqrt{\frac{1}{\left(N_\rmn{elite}-1\right)}\sum\limits_{i=1}^{N_\rmn{elite}}\left[X^\rmn{elite}_{ij}(0)-\bar{p}^\rmn{elite}_j(0)\right]^2}.
\end{equation}

The array ${\bf X}$ at the next iteration is determined as:

\begin{equation}
  X_{ij}(1)=\bar{p}^\rmn{elite}_j(0)+{\bf \sigma}^\rmn{elite}_j(0) G_{ij},
\end{equation}

This process is repeated from equation (2), with $G_{ij}$ regenerated at each iteration. The optimization stops when either the mean value of ${\bf \sigma}^\rmn{elite}_i(k)$ is smaller than a predefined value or the maximum number of iterations $k_\rmn{max}$ is reached.

In order to prevent convergence to a sub-optimal solution due to the intrinsic rapid convergence of the CE method, \citet{kro06} suggested the implementation of a fixed smoothing scheme for $\bar{p}^\rmn{elite,s}_j(k)$ and ${\bf \sigma}^\rmn{elite,s}_j(k)$:

\begin{equation}
  \bar{p}^\rmn{elite,s}_j(k)=\alpha^\prime\bar{p}^\rmn{elite}_j(k)+\left(1-\alpha^\prime\right)\bar{p}^\rmn{elite}_j(k-1),
\end{equation}

\begin{equation}
  {\bf \sigma}^\rmn{elite,s}_j(k)=\alpha_\rmn{d}(k){\bf \sigma}^\rmn{elite}_j(k)+\left[1-\alpha_\rmn{d}(k)\right]{\bf \sigma}^\rmn{elite}_j(k-1),
\end{equation}
where $\alpha^\prime$ is a smoothing constant parameter ($0<\alpha^\prime< 1$) and $\alpha_\rmn{d}(k)$ is a dynamic smoothing parameter at $k$th iteration:

\begin{equation}
  \alpha_\rmn{d}(k)=\alpha-\alpha\left(1-k^{-1}\right)^q,
\end{equation}
with $0<\alpha< 1$ and $q$ is an integer typically between 5 and 10 \citep{kro06}.

As mentioned before, such parametrisation prevents the algorithm from finding a non-global minimum solution since it guarantees polynomial speed of convergence instead of exponential \citep{kro06}.


\begin{table}
 \centering
 \begin{minipage}{80mm}
  \caption{Precession model parameters used to generate artificially the right-ascension and declination offsets of the jet components. The sense of precession was chosen to be clockwise, with the precession period at the observer's reference frame corresponding to 14.20 yr.}
  \begin{tabular}{@{}ccccccc@{}}
  \hline
  Test & $\gamma$ & $\eta_0$ & $\phi_0$ & $\varphi_0$ & $\Delta\tau_\rmn{s}$ & $\mid\Delta_\rmn{noise}\mid$ \\
  & & (deg) & (deg) & (deg) & & (mas) \\
 \hline
 T1 & 8.99 & -161.3 &  11.4 & 8.7 & 3.954 & 0.0\\
 T2 & 8.99 & -161.3 &  11.4 & 8.7 & 3.954 & 0.1\\
 T3 & 8.99 & -161.3 &  11.4 & 8.7 & 3.954 & 0.5\\
 T4 & 10.01 & -165.0 &  15.3 & 5.0 & 3.971 & 0.0\\
 T5 & 7.47 & 118.1 &  4.0 & 0.0 & 3.860 & 0.0\\
\hline
\end{tabular}
\end{minipage}
\end{table}


   \begin{figure}
	  {\includegraphics[width = 86 mm, height = 160 mm]{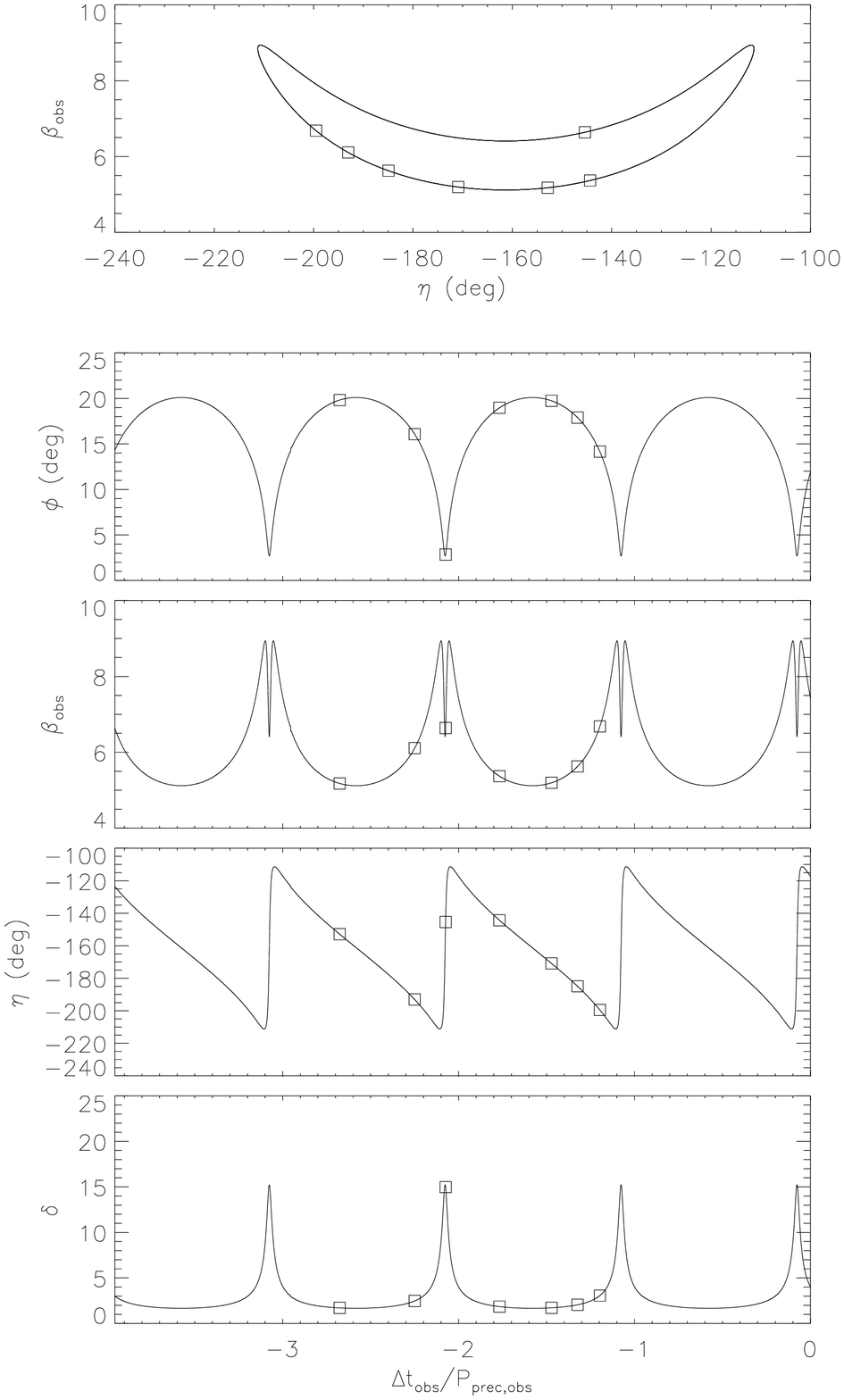}}
      \caption{Jet parameter evolution in the observer's reference frame for the precession model tests T1, T2 and T3. Open squares represent the kinematic parameters used to produce the synthetic components.}
      \label{precession model T1-T3}
   \end{figure}

   \begin{figure}
	  {\includegraphics[width = 86 mm, height = 160 mm]{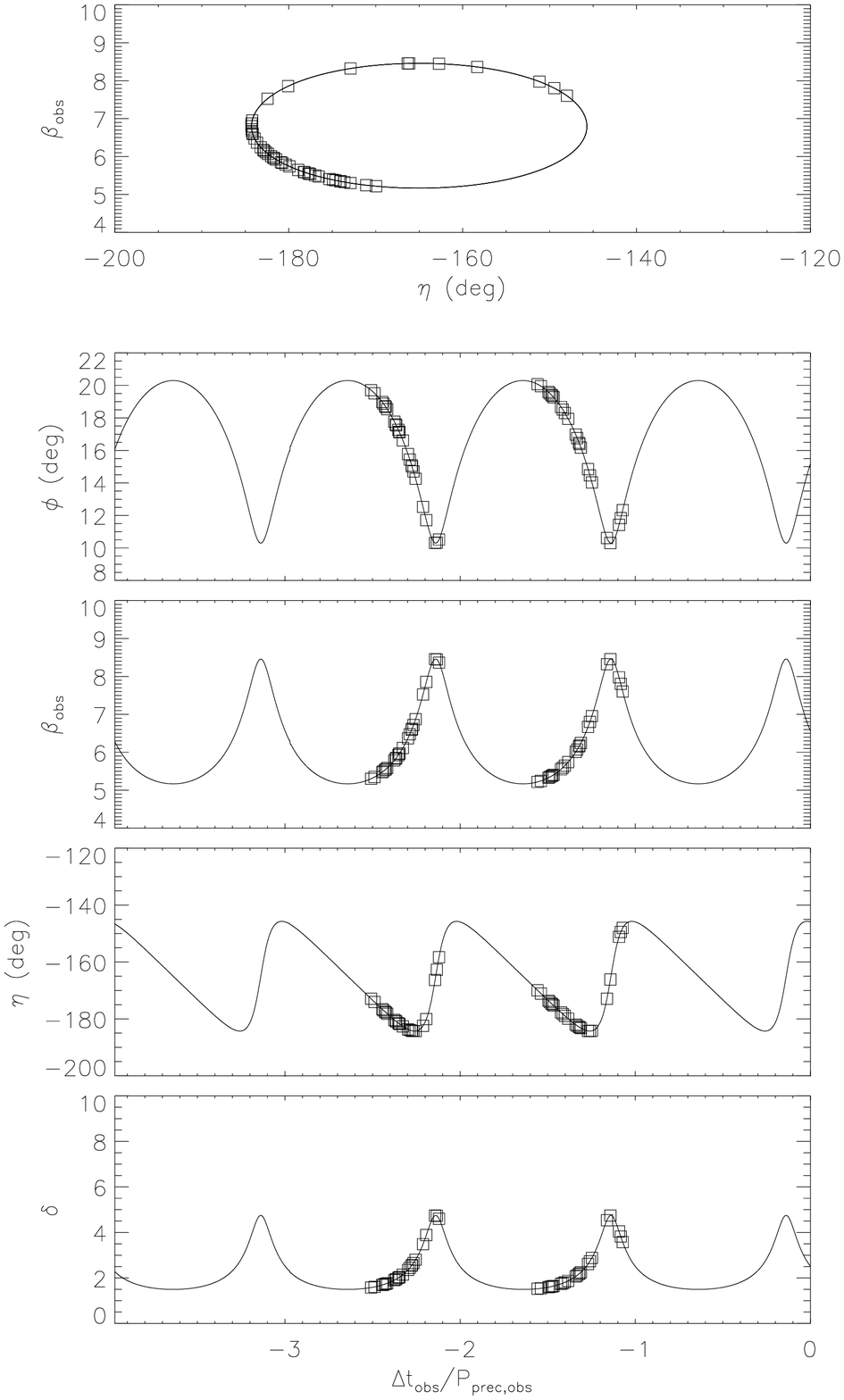}}
      \caption{Jet parameter evolution in the observer's reference frame for the precession model test T4. Open squares represent the kinematic parameters used to produce the synthetic components.}
      \label{precession model T4}
   \end{figure}

   \begin{figure}
	  {\includegraphics[width = 86 mm, height = 160 mm]{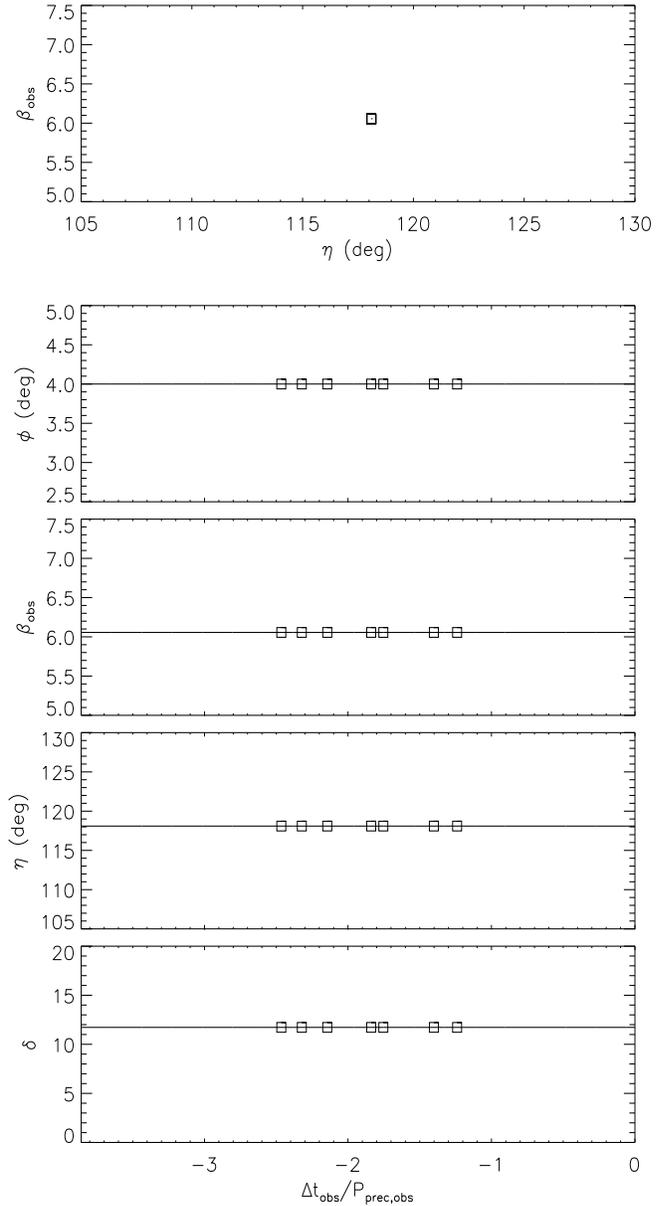}}
      \caption{Jet parameter evolution in the observer's reference frame for the precession model test T5. Open squares represent the kinematic parameters used to produce the synthetic components.}
      \label{precession model T5}
   \end{figure}

\subsection{Ballistic jet precession model}

Let us consider a relativistic jet receding from the core with a constant bulk velocity $\beta$ (in units of the light speed $c$) that precesses around a fixed axis $z_\rmn{s}$, as shown in Fig. 1. Jet and counter-jet recede from the nucleus in opposite directions. Because of precession, the jet inlet direction varies with time with a precession period $P_\rmn{prec,s}$ measured in the source reference 
frame, producing a cone with semi-aperture angle $\varphi_0$. The precession phase 
$\omega_\rmn{s}\Delta t_\rmn{s}=2\pi(t_\rmn{s}-t_\rmn{0,s})/P_\rmn{prec,s}$ is chosen arbitrarily to be 
zero in the $y_\rmn{s}z_\rmn{s}$-plane at $t_\rmn{s}=t_\rmn{0,s}$. From these  definitions, the 
instantaneous jet inlet direction is given in terms of a unit vector with Cartesian components:

 \[
   e_\rmn{x,s}(t_\rmn{s})=\sin\varphi_0\sin[\omega_\rmn{s}(t_\rmn{s}-t_\rmn{0,s})],
 \]
 \[
   e_\rmn{y,s}(t_\rmn{s})=\sin\varphi_0\cos[\omega_\rmn{s}(t_\rmn{s}-t_\rmn{0,s})],
 \]
 \[
   e_\rmn{z,s}(t_\rmn{s})=\cos\varphi_0.  
 \]

Making two consecutive clockwise rotations of the source's coordinate system, the first one by an angle $\phi_0$ around the $y_\rmn{s}$-axis, and the second by an angle $\eta_0$ of the resulting system around the $z$-axis, we have (e.g., \citealt{caab04a,caab04b}):

\begin{equation}
  e_\rmn{x,obs}(t_\rmn{s})=A(t_\rmn{s})\cos\eta_0-e_\rmn{y,s}(t_\rmn{s})\sin\eta_0,
\end{equation}

\begin{equation}
  e_\rmn{y,obs}(t_\rmn{s})=A(t_\rmn{s})\sin\eta_0+e_\rmn{y,s}(t_\rmn{s})\cos\eta_0,
\end{equation}

\begin{equation}
  e_\rmn{z,obs}(t_\rmn{s})=-e_\rmn{x,s}(t_\rmn{s})\sin\phi_0+e_\rmn{z,s}(t_\rmn{s})\cos\phi_0,
\end{equation}
 where:
\begin{equation}
  A(t_\rmn{s})=e_\rmn{x,s}(t_\rmn{s})\cos\phi_0+e_\rmn{z,s}(t_\rmn{s})\sin\phi_0.
\end{equation}

   \begin{figure*}
	  {\includegraphics[width = 140 mm, height = 45 mm]{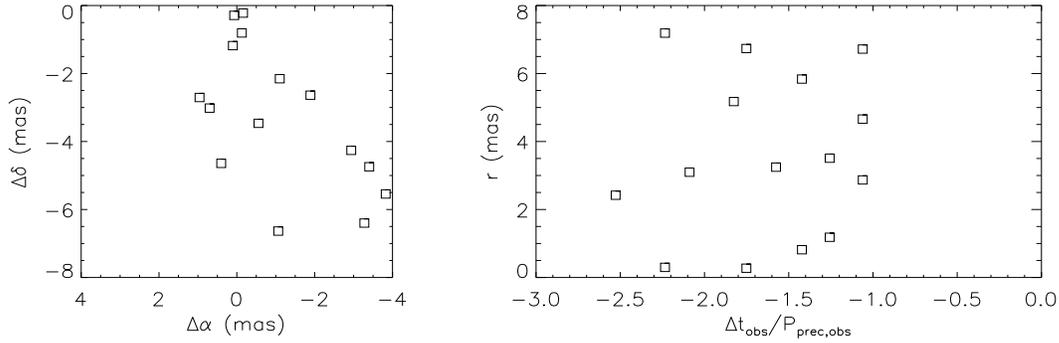}}
      \caption{Right ascension and declination offsets, as well as the core-component distances of the synthetic jet components used in T1-test.}
      \label{radec_T1}
   \end{figure*}

The parameter $\phi_0$  is the angle between the precession cone axis (positive $z_\rmn{s}$-axis direction) and the line of sight (positive $z_\rmn{obs}$-axis direction), and  $\eta_0$ is the position angle of the axis on the plane of the sky (positive from north to east direction).

The instantaneous angle between the jet and the line of sight $\phi(t_\rmn{s})$ is calculated from:

\begin{equation}
  \phi(t_\rmn{s})=\arccos[e_\rmn{z,obs}(t_\rmn{s})],
\end{equation}
 while the position angle of the jet on the plane of the sky is obtained from:
 
\begin{equation}
  \eta(t_\rmn{s})=\arctan\left[\frac{e_\rmn{y,obs}(t_\rmn{s})}{e_\rmn{x,obs}(t_\rmn{s})}\right].
\end{equation}

The observed jet velocity $\beta_\rmn{obs}(t_\rmn{s})$ is:

\begin{equation}
  \beta_\rmn{obs}(t_\rmn{s})=\gamma\beta\delta(t_\rmn{s})\sin\phi(t_\rmn{s}),
\end{equation}
 where the jet Lorentz factor $\gamma$ is  
 
\begin{equation}
  \gamma=\left(1-\beta^2\right)^{-1/2},
\end{equation}
 and the jet Doppler boosting factor $\delta$ is 

 \begin{equation}
  \delta(t_\rmn{s})=\gamma^{-1}\left[1-\beta\cos\phi(t_\rmn{s})\right]^{-1}.
\end{equation}

In order to compare predictions from the precession model with the observational data, it is necessary 
to transform the elapsed time measured in the source's reference frame $dt_\rmn{s}$ to the time interval 
in the observer's framework $dt_\rmn{obs}$. Based on \citet{gow82}, we can write the transformation as:

\begin{equation}   
\frac{\Delta t_\rmn{obs}}{P_\rmn{prec,obs}}=\frac{\int_{0}^{\Delta\tau_\rmn{s}}\delta^{-1}(\tau)d\tau}{\int_0^1\delta^{-1}(\tau)d\tau},
\end{equation}
 where $\Delta t_\rmn{obs}=\left(t_\rmn{obs}-t_\rmn{0,obs}\right)$ and $\Delta\tau_\rmn{s}=\Delta t_\rmn{s}/P_\rmn{prec,s}$.

The relation between the precession period in the source's framework and that measured by the observer $P_\rmn{prec,obs}$ is given as:

\begin{equation}     
   P_\rmn{prec,s}=\frac{\gamma}{(1+z)}\frac{P_\rmn{prec,obs}}{\int_0^1\delta^{-1}(\tau)d\tau},
\end{equation}
where $z$ is the redshift of the source.

   \begin{figure*}
	  {\includegraphics[width = 140 mm, height = 45 mm]{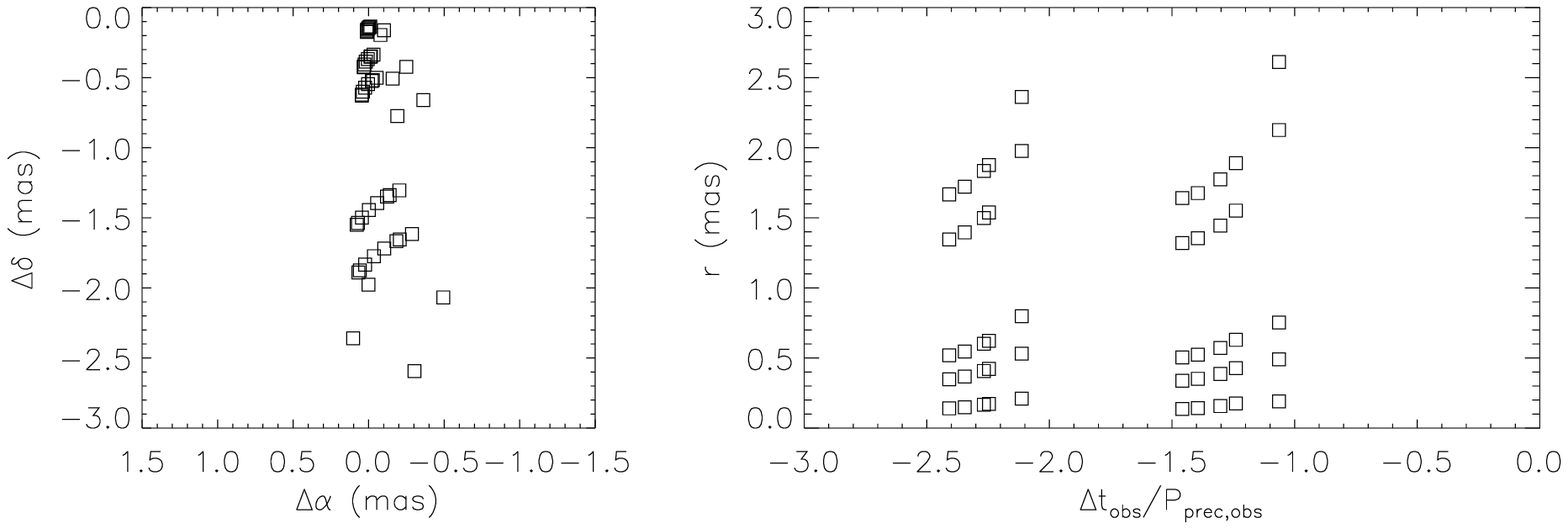}}
      \caption{Right ascension and declination offsets, as well as the core-component distances of the synthetic jet components used in T4-test.}
      \label{radec_T4}
   \end{figure*}

   \begin{figure*}
	  {\includegraphics[width = 140 mm, height = 45 mm]{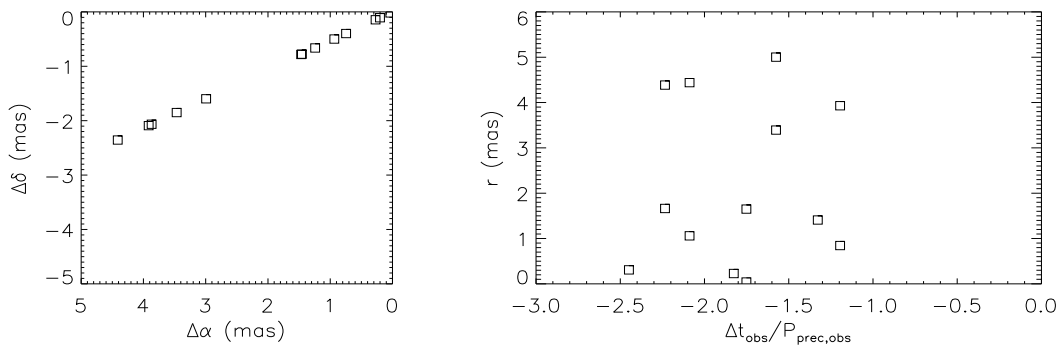}}
      \caption{Right ascension and declination offsets, as well as the core-component distances of the synthetic jet components used in T5-test.}
      \label{radec_T5}
   \end{figure*}

\subsection{Building synthetic jet components}

In order to validate our CE precession model, we constructed five sets of data, simulating right ascension and declination offsets between core and jet components, $\Delta\alpha_\rmn{mod}$ and $\Delta\delta_\rmn{mod}$ respectively. These artificial jet knots were generated from different sets of precession model parameters, as listed in Table 1, assuming that they recede ballistically from the core after their formation. We also imposed arbitrarily a maximum angular separation of about 10 mas to the components. The evolution of $\phi$, $\beta_{\rm obs}$, $\eta$, and $\delta$ in the observer's reference frame, as well as $\beta_{\rm obs}$ vs. $\eta$ for the precession models provided in Table 1 are presented in Figs. 2-4.

Tests T1-T3 consist of a relativistic jet whose precession produces large and rapid time variations in its position angle on the plane of the sky and in its apparent velocity. Changes in these quantities in test T4 are smoother, even though the jet is more relativistic. Test T5 represents the case of a mild relativistic jet with no precession ($\varphi_0 = 0$).

In the case of T1, we generated 15 data points that trace the trajectories of seven components ejected at the epochs shown by the squares in Fig. 2. These data points are sparsely sampled in an interval of about 1.5 precession periods, in order to mimic observations in which the source is not well-sampled in time, as can be seen in Fig.5.

Assuming that the jet was observed at nine different epochs (see Fig. 5), we calculated $\Delta\alpha_\rmn{mod}$ and $\Delta\delta_\rmn{mod}$ from:

\begin{equation}     
   \Delta\alpha_\rmn{mod}(t_\rmn{obs}) = \Delta r_\rmn{mod}(t_\rmn{obs})\sin[\eta_\rmn{mod}(t_\rmn{ej})],
\end{equation} 

\begin{equation}     
   \Delta\delta_\rmn{mod}(t_\rmn{obs}) = \Delta r_\rmn{mod}(t_\rmn{obs})\cos[\eta_\rmn{mod}(t_\rmn{ej})],
\end{equation}
where:

\begin{equation}     
   \Delta r_\rmn{mod}(t_\rmn{obs}) = \mid t_\rmn{obs}-t_\rmn{ej}\mid\mu_\rmn{mod}(t_\rmn{ej}),
\end{equation} 
$\eta_\rmn{mod}$ and $\mu_\rmn{mod}$ are respectively the position angle and the component's proper motion at ejection time $t_\rmn{ej}$.

Tests T2 and T3 are similar to T1 except for the introduction of uniformly-distributed random noise in the right ascension and declination offsets, with maximum amplitude $\Delta_\rmn{noise}$ shown in Table 1. The noise was added to the original coordinates as:

\begin{equation}     
   \Delta\alpha = \Delta\alpha_\rmn{mod}+\Delta^\rmn{RA}_\rmn{noise},
\end{equation}

\begin{equation}     
   \Delta\delta = \Delta\delta_\rmn{mod}+\Delta^\rmn{DEC}_\rmn{noise},
\end{equation}
where the superscripts RA and DEC were used to distinguish the random noise introduced into right ascension and declination respectively. They simulate errors introduced into observational data from the model fitting uncertainties and frequency-dependent shifts in core-component distances produced by opacity effects (e.g., \citealt{blko79,loba98,kov08}). 

The usual procedure before modeling jet precession is to identify the jet knots and follow them in time to obtain their proper motions and position angles on the plane of the sky (e.g., \citealt{abro99,kel04,caab04a,caab04b}). The lack of close enough successive observations or high rates of jet component production might contribute to an ambiguous component identification, as well as its accurate kinematic behaviour. 
 
Motivated by this, we generate randomly 50 components spread out in an interval of about 1.3 precession periods, as shown in Fig. 6. Each component is unique (one data point for each jet component) and obeys kinetically the T4 precession model. Note that the plot of core-component distance as a function of time in Fig. 6 seems to suggest (wrongly!) the existence of about seven components ejected with low velocities and receding non-ballistically from the core.

Test T5 consists of a non-precessing jet, in which the components move at a position angle of 118$\fdg$1 with constant velocity ($\beta_\rmn{obs}\sim 6.1$). We created 13 data points, corresponding to seven jet components launched at the epochs represented in Fig. 4. Plots of right ascension and declination offsets, as well as those for the core-component distances for T5, can be seen in Fig. 7.

\subsection{Estimating precession parameters from cross-entropy method}

All validation tests were performed with a fixed $P_\rmn{prec,obs}$, reducing to five the free model parameters to be optimized by the CE method ($\gamma, \eta_0, \phi_0, \varphi_0$ and $\Delta\tau_\rmn{s}$)\footnote{The inclusion of $P_\rmn{prec,obs}$ as an additional parameter to be optimised brings the necessity to parallelize the algorithm in order to avoid very long processing time (the IDL version of the code used in this work spends about 20 hours to generate a particular entry in Table 2 in a 2.4 GHZ Intel(R) Core(TM) 2 Quad processor). This will be implemented in future.}. The reason for this is that precession period is usually constrained by other studies, such as periodic variations in the continuum spectrum (e.g., \citealt{gia73,leva96,caab04a,caab04b}) and in velocity or/and intensity of emission lines (e.g. \citealt{marg84,sti02,sto03}). 

In order to obtain the optimal precession parameters through the CE optimization method, the values of $N$, $\alpha$, $N_\rmn{elite}$ and $q$ must be provided. We decided to use $q=5$ for all tests in Table 1 in order to guarantee a polynomial speed of convergence, as mentioned in Section 2 (see section 3.2 for some consequences of departures from that value). For the other parameters, we have used $10\leq N\leq 100$, $0.6\leq\alpha\leq 0.9$ and $5\leq N_\rmn{elite}\leq 10$. We also adopted  $\alpha^\prime=1$, which makes equation (5) not critical to this work ($\bar{p}^\rmn{elite,s}_j(k)=\bar{p}^\rmn{elite}_j(k)$).

At each iteration $N$  sets of random  precession parameters are generated, based on equations (4) and (5), which are used to determine $\beta_\rmn{obs}$, $\eta_\rmn{obs}$, $\delta_\rmn{obs}$ and $\Delta t_\rmn{obs}$ from equations (12)-(17). Then $\Delta\alpha_\rmn{mod}$, $\Delta\delta_\rmn{mod}$ and $\Delta r_\rmn{mod}$ are calculated using equations (19)-(21) and compared quantitatively with the data through the objective function. The elite sample, characterized by the $N_\rmn{elite}$-lowest values of $S$, is chosen to be used as input into equations (4) and (5) to produce the next set of precession parameters.


\begin{table*}
 \centering
 \begin{minipage}{127mm}
  \caption{Results for the jet precession model parameters from cross-entropy method for $q=5$ and $k_\rmn{max}=500$. The columns 5-9 show the relative errors among the real precession model parameters and their three-run mean values obtained from the CE optimisation (subscripts refer to the corresponding precession parameters).}
  \begin{tabular}{@{}cccccccccc@{}}
  \hline
  Test & $N$ & $N_\rmn{elite}$ & $\alpha$ & $\epsilon_\gamma$ & $\epsilon_{\eta 0}$ & $\epsilon_{\phi 0}$ & $\epsilon_{\varphi 0}$ & $\epsilon_{\Delta\tau\rmn{s}}$ & $S(k_\rmn{max})$\\
  & & & & ($\%$) & ($\%$) & ($\%$) & ($\%$) & ($\%$) &\\
 \hline
 T1 & 10  & 5  & 0.6 & 67.79      &      19.76      &     288.71      &     335.25      &     0.1330      &      5.161  $\pm$      0.190\\
 T1 & 20  & 5  & 0.6 & 33.59      &       9.80      &     144.29      &     168.02      &     0.1139      &      2.327  $\pm$      0.032\\
 T1 & 30  & 5  & 0.6 & 14.19      &      21.53      &     315.91      &     359.77      &     0.4984      &      4.926  $\pm$      0.127\\
 T1 & 40  & 5  & 0.6 &  7.09      &       1.24      &       6.74      &      11.65      &     0.0044      &      1.151  $\pm$      0.080\\
 T1 & 50  & 5  & 0.6 &  4.81      &       0.24      &       0.42      &       3.20      &     0.0020      &      0.963  $\pm$      0.091\\
 T1 & 70  & 5  & 0.6 &  1.79      &       0.07      &       0.05      &       1.27      &     0.0089      &      0.209  $\pm$      0.000\\
 T1 & 100 & 5  & 0.6 &  7.32      &       0.26      &       0.41      &       3.75      &     0.0046      &      0.265  $\pm$      0.011\\
 T1 & 50  & 5  & 0.7 &  1.00      &       0.00      &       0.36      &       0.26      &     0.0011      &      0.943  $\pm$      0.091\\
 T1 & 50  & 5  & 0.9 &  1.45      &       0.01      &       0.34      &       0.68      &     0.0019      &      0.900  $\pm$      0.098\\
 T1 & 50  & 10 & 0.6 & 10.43      &       0.01      &       3.35      &       4.05      &     0.0582      &      3.942  $\pm$      0.098\\
 T1 & 50  & 10 & 0.7 &  4.77      &       3.14      &      30.92      &      42.46      &     0.0300      &      5.166  $\pm$      0.382\\
 T1 & 50  & 10 & 0.9 &  2.61      &       0.10      &       1.30      &       0.18      &     0.0004      &      2.794  $\pm$      0.179\\
 T2 & 50  & 5  & 0.6 &  4.65      &       0.34      &       8.93      &       7.19      &     0.1675      &      1.674  $\pm$      0.075\\
 T2 & 70  & 5  & 0.6 &  1.74      &       0.38      &       2.89      &       4.14      &     0.2029      &      0.147  $\pm$      0.002\\
 T2 & 100 & 5  & 0.6 &  3.21      &       0.28      &       2.79      &       3.08      &     0.2211      &      2.837  $\pm$      0.190\\
 T2 & 50  & 5  & 0.7 &  8.79      &       0.45      &       1.11      &       2.27      &     0.2392      &      0.258  $\pm$      0.020\\
 T2 & 50  & 5  & 0.9 &  2.99      &       0.28      &       2.40      &       2.63      &     0.2052      &      1.592  $\pm$      0.083\\
 T2 & 50  & 10 & 0.6 & 69.11      &      20.70      &     306.18      &     354.50      &     0.3292      &      4.957  $\pm$      0.281\\
 T2 & 50  & 10 & 0.7 & 43.85      &      12.31      &     183.77      &     212.87      &     0.0101      &      2.254  $\pm$      0.012\\
 T2 & 50  & 10 & 0.9 &  1.79      &       0.34      &       1.83      &       5.39      &     0.2259      &      0.869  $\pm$      0.081\\
 T3 & 50  & 5  & 0.6 & 31.76      &       6.41      &      71.38      &      61.35      &     0.7650      &      3.643  $\pm$      0.069\\
 T3 & 70  & 5  & 0.6 & 23.68      &       3.51      &      26.35      &       5.05      &     0.7450      &      3.880  $\pm$      0.062\\
 T3 & 100 & 5  & 0.6 & 23.77      &       3.56      &      26.64      &       5.31      &     0.7428      &      3.442  $\pm$      0.065\\
 T4 & 50  & 5  & 0.6 &  0.96      &       0.02      &       0.87      &       0.59      &     0.0086      &   0.000236  $\pm$   0.000001\\
 T4 & 70  & 5  & 0.6 &  0.10      &       0.01      &       0.51      &       0.36      &     0.0070      &   0.000182  $\pm$   0.000001\\
 T4 & 100 & 5  & 0.6 &  0.76      &       0.01      &       0.31      &       0.23      &     0.0033      &   0.000166  $\pm$   0.000000\\
 T4 & 50  & 5  & 0.7 &  0.29      &       0.00      &       0.16      &       0.20      &     0.0032      &   0.000136  $\pm$   0.000000\\
 T4 & 50  & 5  & 0.9 &  1.03      &       0.04      &       1.93      &       1.48      &     0.0242      &   0.400128  $\pm$   0.043864\\
 T4 & 50  & 10 & 0.6 &  2.37      &       0.05      &       1.41      &       0.96      &     0.0134      &   0.001663  $\pm$   0.000026\\
 T4 & 50  & 10 & 0.7 &  1.01      &       0.03      &       0.72      &       0.46      &     0.0062      &   0.000455  $\pm$   0.000006\\
 T4 & 50  & 10 & 0.9 &  0.81      &       0.03      &       1.39      &       1.08      &     0.0181      &   0.000152  $\pm$   0.000000\\
 T5 & 50  & 5  & 0.6 & 37.70      &       0.00      &    1143.97      &       0.00      &     0.6123      &   0.000003  $\pm$   0.000000\\
 T5 & 70  & 5  & 0.6 & 68.64      &       0.00      &    1139.88      &       0.00      &     0.3440      &   0.000002  $\pm$   0.000000\\
 T5 & 100 & 5  & 0.6 & 11.94      &       0.00      &    1129.21      &       0.00      &     0.9651      &   0.000003  $\pm$   0.000000\\
\hline
\end{tabular}
\end{minipage}
\end{table*}


The objective function $S(k)$ transmits to the CE algorithm the best tentative solutions at the $k$th iteration. We have chosen $S(k)$ as:

\begin{equation}     
   S(k)= \Upsilon(k)+\sum\limits_{i=1}^{N_\rmn{d}}\left\{\left[S_{\alpha_i}(k)\right]^2+\left[S_{\delta_i}(k)\right]^2+\left[S_{r_i}(k)\right]^2\right\},
\end{equation} 
where:

\begin{equation}     
   S_{\alpha_i}(k)= \Delta\alpha_i-\Delta\alpha_\rmn{mod_i}(k),
\end{equation}

\begin{equation}     
   S_{\delta_i}(k)= \Delta\delta_i-\Delta\delta_\rmn{mod_i}(k),
\end{equation}

\begin{equation}     
   S_{r_i}(k)= \Delta r_i-\Delta r_\rmn{mod_i}(k),
\end{equation}
with $\Delta r_i^2=\Delta\alpha_i^2+\Delta\delta_i^2$. It is important to emphasise that the terms $S_{\alpha_i}$ and $S_{\delta_i}$ strongly constrain the instantaneous jet direction in the optimization process, while inclusion of $S_{r_i}$ provides additional constraint to the modelling of the core-component distances besides improving the convergence performance of the method. The penalty function $\Upsilon(k)$ is given as:

\begin{equation}     
  \Upsilon(k) = \left\{\begin{array}{rc}
0&\mbox{if}\quad \mid t_{\zeta \delta_\rmn{max}}(k)-t_\rmn{peak}\mid\le \epsilon\\
10 &\mbox{if}\quad \mid t_{\zeta \delta_\rmn{max}}(k)-t_\rmn{peak}\mid\ge \epsilon
\end{array}\right.
\end{equation}
where $t_\rmn{peak}$ is the epoch at which there is an outburst in the light curve, which we associate  with the radiation boosting of the underlying jet (see e.g. \citealt{caab04b} for more details), $t_{\zeta \delta_\rmn{max}}$ is the epoch at which the Doppler boosting factor $\delta$ reaches  $\zeta$ times its maximum value ($0\le\zeta\le 1$) and $\epsilon$ is a constant interpreted as the uncertainty in the epoch of occurrence of the maximum of the jet radiation boosting in the light curve. In this work, we adopted  $\zeta=0.9$ and $\epsilon=4\%$ of the precession  period.

We can see that the choice of $S$ is based on the minimisation of the quadratic distances between the observational data and those produced by the precession model, while $\Upsilon(k)$ provides an extra constraint in the $\Delta\tau_\rmn{s}$ parameter. It is important to emphasise that the CE method allows the inclusion on the calculations of as many penalty functions as necessary to improve the convergence of the method.

The search for the optimal parameters was halted at iteration $k_\rmn{max}=500$ for all tests listed in Table 1. Good estimates of the precession parameters were obtained at $k_\rmn{max}$ in most cases (see next section). An additional motivation to limit the number of iterations was checking the convergence's behaviour of the CE method as $\alpha$ and $N_\rmn{elite}$ have their values varied. 

Each validation test was performed three times, having the CE values for $\gamma, \eta_0, \phi_0, \varphi_0$ and $\Delta\tau_\rmn{s}$ been calculated averaging their $k_\rmn{max}$-values weighted by the inverse of $S^2(k_\rmn{max})$, where $S$ is averaged over $N_\rmn{elite}$-best solutions at $k_\rmn{max}$. Their respective uncertainties were obtained from the calculation of their standard deviations.

Note that all validation tests presented in this work do not use any information concerning counter-jet data. In the case of micro-quasars and Herbig-Haro sources, the inclusion of counter-jet data must increase the performance of the CE optimisation. The reason for this is that the presence of the counter-jet in the images imposes additional constraint on the Doppler boosting parameter, which on the other hand depends on the jet speed and the jet viewing angle. On the other hand, the relativistic counter-jets in BL Lac objects and quasars are significantly Doppler-deboosted for small viewing angles, which difficult their direct detections, as well as any kinematic study of their knots. For radio galaxies, the long timescales involved on the changes of the jet/counter-jet structures limit the detectability of the precession in terms of the component motions. Nevertheless, if the jet-counter-jet intensity ratio is known, it can be used in all cases above to put additional constraints on the value of Doppler-boosting factor during the optimisation process.

   \begin{figure*}
	  {\includegraphics[width = 175 mm, height = 160 mm]{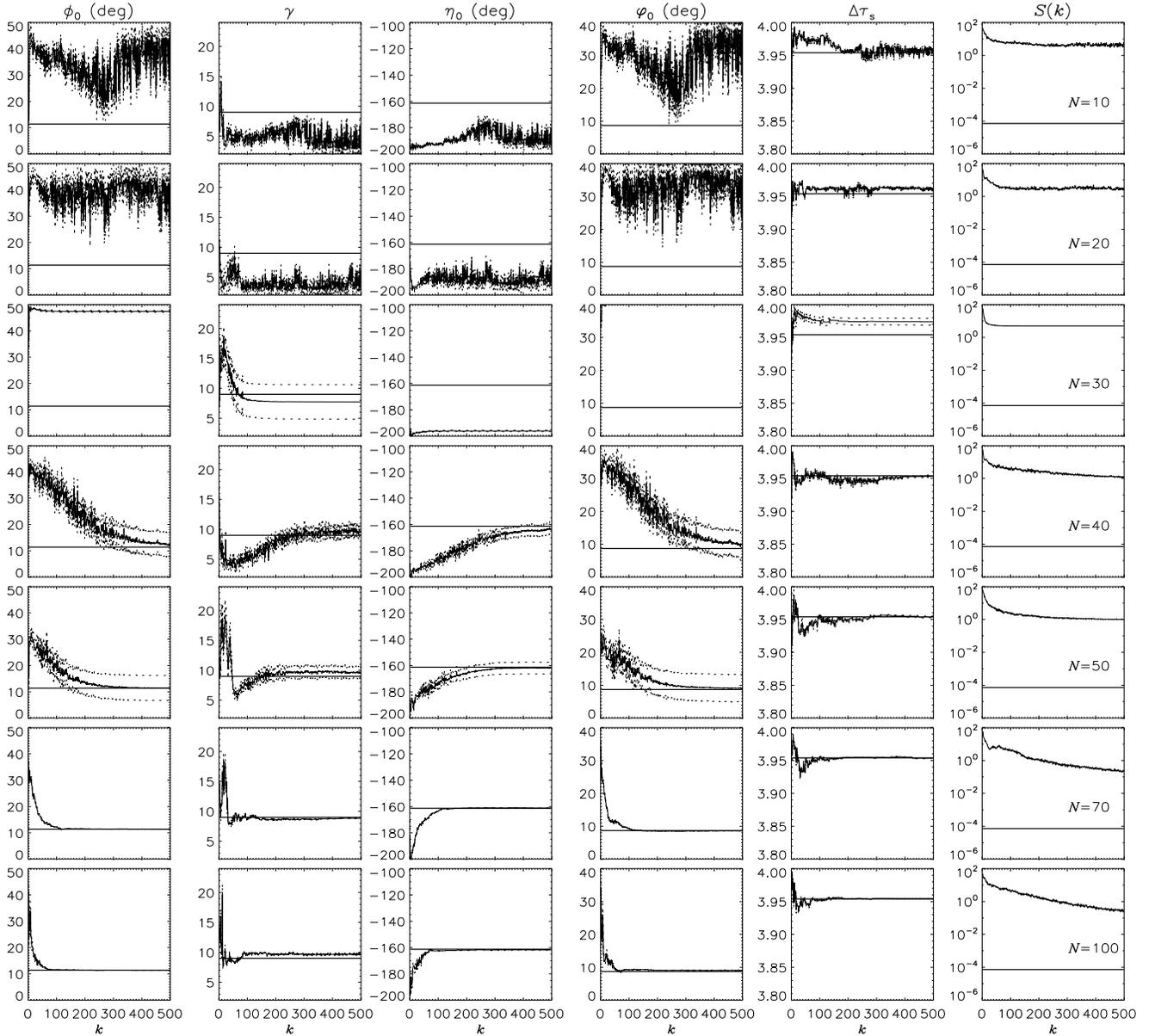}}
      \caption{Evolution of the precession parameters (first five columns from left to right) as well as the value of the objective function (last column) for T1 (see Table 1) as a function of the iteration number $k$, assuming $N_\rmn{elite}=5$, $\alpha=0.6$ and $q=5$. From top to bottom we have $N=10, 20, 30, 40, 50, 70$ and 100 respectively (small labels were put into $S$-plots to guide the reader). Thick solid lines show the real precession model parameters. The mean parameter values (weighted by the squared inverse values of the objective function) recovered by the cross-entropy algorithm after three independent runs, as well as their $3\sigma$-deviations, are displayed by thin solid and dotted lines respectively.}
      \label{parameter evolution N}
   \end{figure*}

   \begin{figure*}
	  {\includegraphics[width = 175 mm, height = 140 mm]{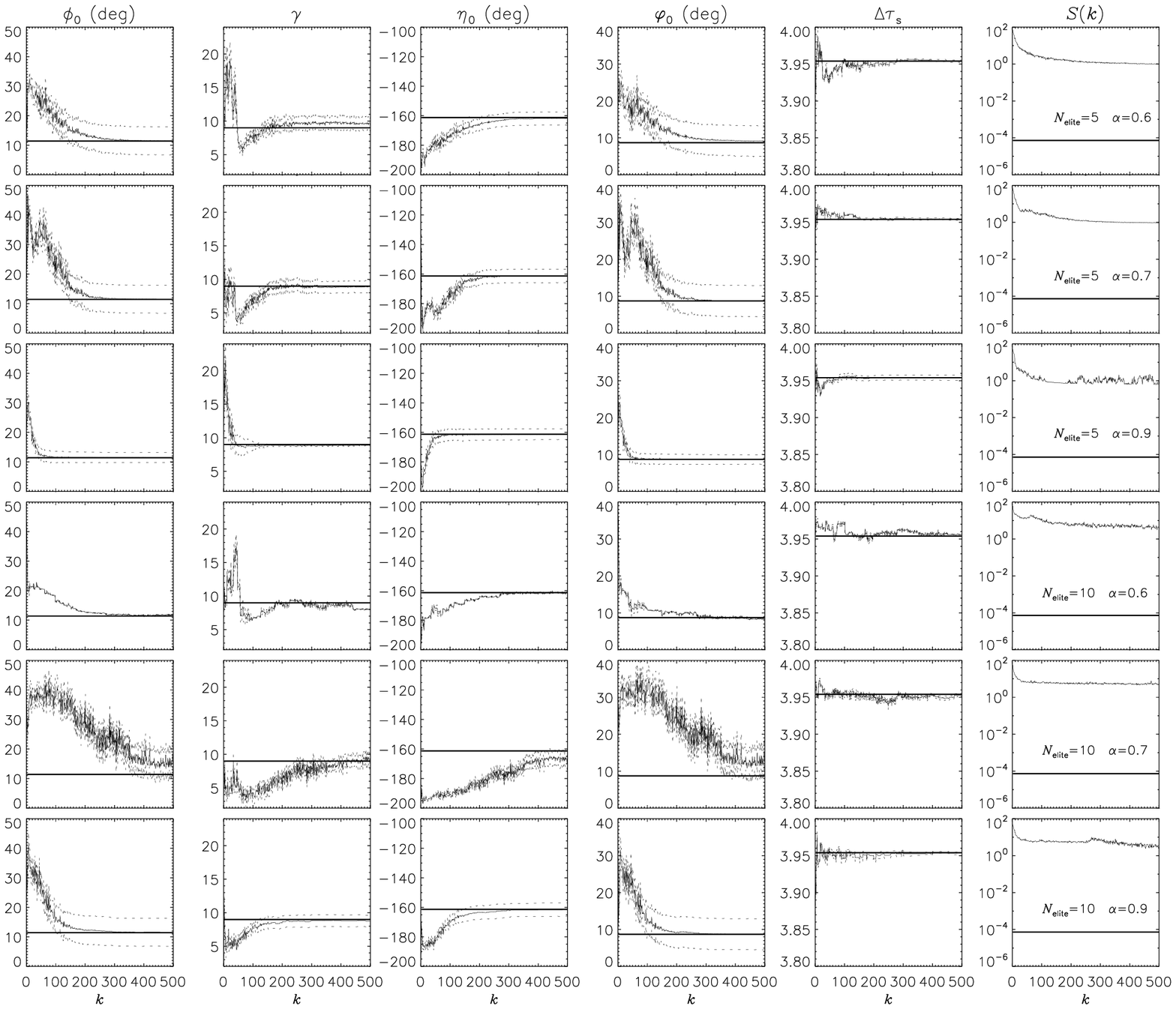}}
      \caption{Evolution of the precession parameters (first five columns from left to right) as well as the value of the objective function (last column) for T1 (see Table 1) as a function of the iteration number $k$, assuming $N=50$ and $q=5$. From top to bottom, the first set of graphs correspond to $N_\rmn{elite}=5$ and $\alpha = 0.6, 0.7$ and 0.9 respectively, while for the last set $N_\rmn{elite}=10$ and $\alpha = 0.6, 0.7$ and 0.9 (small labels were put into $S$-plots to guide the reader). Thick solid lines show the real precession model parameters. The mean parameter values (weighed by the squared inverse values of the objective function) recovered by the cross-entropy algorithm after three independent runs, as well as their $3\sigma$-deviations, are displayed by thin solid and dotted lines respectively.}
      \label{parameter evolution alpha}
   \end{figure*}

\section{Results and discussion}

\subsection{Tests T1-T5}
There is not yet a theoretical foundation that guarantees which CE parameters must be used in the calculations in order to maximise the optimisation efficiency (e.g., \citealt{kro06}). Therefore, it is necessary to explore the CE parameter-space in order to estimate the optimal or near-optimal values.

We present in Table 2 the relative difference between the precession model parameters and those derived from the CE technique for all tests listed in Table 1, as well as the CE parameters $N$, $N_\rmn{elite}$, and $\alpha$ used in the calculations. All runs were arbitrarily halted at iteration $k_\rmn{max}=500$, which means that not necessarily the optimisations  reached their best value in all cases. Nevertheless, the results presented here are representative of the great potential of the cross-entropy continuous optimization technique as a quantitative tool for jet precession inference.

\subsubsection{T1}

The first seven runs of T1 were made fixing $N_\rmn{elite}=5$ and $\alpha=0.6$ varying  $N$ from 10 to 100. The precession parameters were allowed to vary in a wide  range : $2.3\leq\gamma\leq 22.4$, $-200\degr\leq\eta_0\leq -100\degr$, $0\fdg 1\leq\phi_0\leq 50\degr$, $0\degr\leq\varphi_0\leq 40\degr$ and $3.8\leq\Delta\tau_\rmn{s}\leq 4.0$. The results are shown in Fig. 8, in which  the evolution of the precession parameters and the objective function are plotted as a function of the iteration number. 

We can see that the increase of $N$ results in a better convergence to the correct precession parameters. Indeed, only  runs with $N\ga 50$ were successful in recovering the model parameters within 3$\sigma$ (or 5$\%$ in terms of relative error). For $N < 50$, the capability of finding the right solution was substantially depreciated due to either the large  variation of the parameters within the three-run optimization  or the rapid convergence to sub-optimal parameters (e.g. see the evolution of $\phi_0$ for $N=30$ in Fig. 8). 

In a general context, \citet{deb05} and \citet{kro06} adopted respectively $N=(5-10)N^2_\rmn{p}$ and $N=100N_\rmn{p}$ in their calculations, which in our case corresponds to $125\leq N\leq 250$ and $N=500$, respectively. Our results  suggest the existence of a minimum value for the sample size $N$ for which the CE optimization can be applied to find the jet precession model parameters\footnote{It is valid for all validation tests presented in this work.}: $N\ga 10N_\rmn{p}$. 

We carried out five additional runs for T1 fixing $N=50$ and assuming $N_\rmn{elite}=5$ and 10, as well as $\alpha=0.6$, 0.7 and 0.9, as shown in Table 2. The performance of the CE optimization is presented in Fig. 9. As expected, the increase of $\alpha$ provides a faster convergence since the mix between consecutive iterations is diminished and the parameters always converged to the real model values, while  $N_\rmn{elite}=10$ increased the  convergence time. The best performance was achieved for $N_\rmn{elite}=5$ and $\alpha=0.9$ for the tests shown in Fig. 9; however we adopted the more conservative value $\alpha=0.6$ in the remaining tests.

\subsubsection{T2}

T2 validation tests are similar to those of T1, except for the introduction of random noise with maximum amplitude of 0.1 mas. The objective was to analyse the CE optimization performance in a more realistic situation. 

First, we checked the influence of $N$ in the optimization process, fixing $N_\rmn{elite}=5$ and $\alpha=0.6$ and assuming $N=50$, 70 and 100. The results are shown in Table 2. Although $N=70$ has minimised slightly better the mean value of $S$, the estimates of the precession parameters for $N=100$ are closer to the real ones ($< 3\%$). 

We also carried out five additional runs for T2 fixing $N=50$ and assuming $N_\rmn{elite}=5$ and 10, as well as $\alpha=0.6$, 0.7 and 0.9. The results are shown in Table 2. Analogously to T1, the best performance has been achieved for $N_\rmn{elite}=5$ and $\alpha=0.9$.

\subsubsection{T3}

Here we added to T1 data random noise with maximum amplitude of 0.5 mas. This is an extreme validation test since uncertainties in right ascension and declination offsets are often smaller, especially at high-frequency observations ($\ga 22$ GHz), and at core-component distances smaller than about 3 mas. The CE optimization behaviour for T3, using $N_\rmn{elite}=5$ and $\alpha=0.6$ and assuming $N=50$, 70 and 100 is presented in Table 2. 

The CE technique for $N=50$ recovered the precession parameters with less efficiency than for $N=70$ and 100, reaching relative errors of about $71\%$ in $\phi_0$, $61\%$ in $\varphi_0$ and $32\%$ in $\gamma$. There were no substantial differences in the precession parameters values obtained from $N=70$ and 100 ($<0.3\%$), with the relative error smaller than $\sim 27\%$ in the worst case.

\subsubsection{T4}

Test T4 was thought to be representative of those observations in which the jet components are not well-identified to be followed along the different epochs of observation. As discussed previously, each one of the 50 points in T4 corresponds to a different jet component. 

This challenging test was conducted maintaining $N_\rmn{elite}=5$ and $\alpha=0.6$ fixed and assuming $N=50$, 70 and 100 in order to study the influence of $N$ in the optimization process. The results are listed in Table 2. We can see that the precession parameters were recovered in all cases, with a relative error smaller than $1\%$. We believe that one of the reasons behind this impressive result is the substantial increase of the number of the jet components used to constrain the CE optimization in contrast to previous tests (50 against 15 data points in T1-T3).

A similar good performance was obtained for the five additional runs for T4, now fixing $N=50$ and assuming $N_\rmn{elite}=5$ and 10, as well as $\alpha=0.6$, 0.7 and 0.9. The differences in the final precession parameters from those tests were no larger $2\%$, having a slightly better performance been observed for $N_\rmn{elite}=5$ and $\alpha=0.7$, as it can be seen in Table 2.

It is important to emphasise that such impressive performance was obtained using data points without any deviation from the pure predictions from the T4 precession model. In practice, jet components could present more complex motions due to e.g. jet-environment interaction, hydro- and magnetohydrodynamical instabilities, as well as core-component distance shifts introduced by opacity effects, which would decrease the recovering accuracy of the CE precession method. Indeed, we made an extra test, adding random noise with maximum amplitude of 0.1 mas in the right ascension and declination offsets in the 50 data points of test T4, assuming $N=100$, $N_\rmn{elite}=5$ and $\alpha=0.6$. As expected, there was a substantial reduction in the algorithm performance, which recovered the real precession parameters within a mean relative error of about 13$\%$. However, more realistic errors (e.g. smaller for shorter distances to the core) are expected to increase the algorithm performance.

\subsubsection{T5}

What would be the  behaviour of the CE algorithm if there were no jet precession? Would the algorithm be able to provide any information about this non-precessing jet? To address these questions we performed test T5, a non-precessing mildly relativistic jet model with $\gamma = 7.5$ (see Table 1 for more details).

The results from this test are presented in Table 2. The parameters $\eta_0$ and $\varphi_0$ were recovered by the optimization process strikingly well. The relative errors in $\gamma$ were about $38\%$, $68\%$ and $12\%$ for $N=50, 70$ and 100 respectively, while for $\Delta\tau_\rmn{s}$ they were always smaller than 1$\%$. However the CE optimization was not able to find the correct value of $\phi_0$. It is important to emphasise that we do not expect to recover all parameters in this situation since we are applying a jet precession model to a non-precessing jet, which obviously can bring some difficulty to the algorithm.

We believe that the failure in recovering the parameters $\phi_0$ and $\gamma$ was caused by degenerate solutions when the jet viewing angle is constant in time. If $\varphi_0=0$, $\mu_\rmn{mod}$ and $\eta_\rmn{mod}$ in equations (19)-(21) are time independent, so that for a given value of $t_\rmn{ej}$ it is possible to find $\mu_\rmn{mod}$ (that depends on $\gamma$ and $\phi_0$) that provide an acceptable solution in the right ascension-declination plane. For example, if we decrease $|t_\rmn{obs}-t_\rmn{ej}|$ by a factor two and increase $\mu_\rmn{mod}$ by the same factor, the displacement of the component remains completely unchanged. An alternative to break this degeneracy would be to introduce an additional constraint on Doppler boosting factor (e.g., using some value for Doppler factor derived from high-energy observations), which means further constraints to $\gamma$ and $\phi_0$. Another possibility could be the usage of the CE method to verify whether there is any signature of jet precession (from whether or not $\varphi_0=0$). If not, then we could return to the standard procedures based on some identification component scheme, such as the time evolution of the jet knots in a plot of core-component distance, which constrains their apparent proper motions and hence the values of $\gamma$ and $\phi_0$.

Nevertheless, it is important to emphasise that the CE precession method was capable to find easily the right zero-value for $\varphi_0$, which strongly suggests its usage to verify the existence of any precession signature in the observational data.

\subsection{Additional tests}

We now discuss additional validation tests concerning changes in the values of the CE parameter $q$ and of the precession period in the observer's reference frame $P_\rmn{prec,obs}$. The influence of the sense of jet precession on the optimization process will be also discussed in this section.


\begin{table}
 \centering
  \caption{Jet precession model parameters from cross-entropy method for T2-test considering additional values of $q$. The CE parameters used in the calculations were $N=50$, $N_\rmn{elite}=5$ and $k_\rmn{max}=500$.}
  \begin{tabular}{@{}cccccccc@{}}
  \hline
  $q$ & $\alpha$ & $\epsilon_\gamma$ & $\epsilon_{\eta 0}$ & $\epsilon_{\phi 0}$ & $\epsilon_{\varphi 0}$ & $\epsilon_{\Delta\tau\rmn{s}}$ & $S(k_\rmn{max})$\\
  & & ($\%$) & ($\%$) & ($\%$) & ($\%$) & ($\%$) & \\
 \hline
   1 & 0.6 & 69.3      &      18.4      &     286.4      &     332.0      &     0.8   &     15.91  $\pm$      0.56\\
   1 & 0.7 & 73.2      &      19.9      &     301.1      &     349.4      &     0.8   &     16.07  $\pm$      0.90\\
   1 & 0.9 & 64.1      &      16.5      &     252.4      &     289.4      &     0.7   &      9.58  $\pm$      0.14\\
  10 & 0.6 & 68.6      &      17.0      &     242.7      &     283.9        &    0.3   &      2.71  $\pm$      0.04\\
  10 & 0.7 & 51.5      &      12.1      &    166.0       &     191.0        &    0.3   &     4.41  $\pm$      0.21\\
  10 & 0.9 & 64.9      &      16.9      &    256.6       &     294.4        &    0.6   &     2.77  $\pm$      0.04\\
\hline
\end{tabular}
\end{table}


\subsubsection{Dynamical smoothing}

All validation tests discussed in last section were done assuming $q=5$. Roughly speaking, this parameter is responsible for controlling the speed of convergence of the CE optimization \citep{kro06}. Increasing its value makes convergence faster, even though it might lead to a sub-optimal solution. On the other hand, low values of $q$ imply in a slower convergence, requiring higher number of iterations  to obtain good solutions.

The precession parameters  for additional runs using $q=1$ and 10 with the T2 data are listed in Table 3. For each of them, we also varied $\alpha$ to probe possible differences in the optimization performance. As expected, the adoption of $q=1$ and 10 worsened substantially the performance of the CE method, so that none of the precession parameters were correctly recovered. Furthermore, we can note that $S(k_\rmn{max})$ listed in Table 3 are higher than those obtained in the same circumstances assuming $q=5$ (see Table 2). 

Therefore, it seems that $q\approx 5$ makes the CE optimization more stable and efficient at least in the case of our jet precession model.


\begin{table}
 \centering
  \caption{Jet precession model parameters from cross-entropy method for T2-test considering additional values for the precession period ($f$ is given in units of $P_\rmn{prec,obs}$ used to build tests T1-T5). The CE parameters used in the calculations were $N=50$, $N_\rmn{elite}=5$, $\alpha=0.6$, $q=5$ and $k_\rmn{max}=500$.}
  \begin{tabular}{@{}ccccccc@{}}
  \hline
  $f$ & $\epsilon_\gamma$ & $\epsilon_{\eta 0}$ & $\epsilon_{\phi 0}$ & $\epsilon_{\varphi 0}$ & $\epsilon_{\Delta\tau\rmn{s}}$ & $S(k_\rmn{max})$\\
   & $(\%)$ & $(\%)$ & $(\%)$ & $(\%)$ & $(\%)$ & \\
 \hline
  0.5  & 70.7      &      11.3      &       5.2      &     342.6      &   102.5      &      1.97  $\pm$      0.01\\
 0.7  & 72.0      &      13.7      &     263.5      &     326.2      &    38.8      &      8.19  $\pm$      0.37\\
 1.0  & 4.65      &       0.34      &       8.93      &       7.19      &     0.1675      &      1.674  $\pm$      0.075\\
 1.4  & 75.9   &   7.8   &     180.2   &     119.0   &    32.9   &    13.37  $\pm$   0.04\\
 2.0  & 148.5      &       0.4      &     154.8      &      15.9      &    64.4      &     14.16  $\pm$      0.00\\
\hline
\end{tabular}
\end{table}


\subsubsection{Sense of precession}

Tests T1-T4 were constructed adopting clockwise precession in the models and in the optimisation procedure. The question that arises is what would happened if we had chosen the wrong sense of precession? 

In order to provide some quantitative answer to this question we used the T2-test data considering the opposite direction for jet precession $(\omega_\rmn{s}<0)$. The CE parameters used in the calculations were $N=50$, $N_\rmn{elite}=5$, $\alpha=0.6$, $q=5$ and $k_\rmn{max}=500$, which led to $\gamma= 2.69 \pm 0.03$, $\eta_0=-179.83\pm 1.28$, $\phi_0=20.60\pm 1.90$, $\varphi_0=14.12\pm 1.39$, $\Delta\tau_\rmn{s}=3.9288 \pm 0.0033$ and $S(k_\rmn{max})=6.907\pm 0.294$. The counterclockwise values are far from the real ones and they also produce a higher $S(k_\rmn{max})$ in comparison to the clockwise precession. Therefore, our CE precession model can also indicate the correct sense of the jet precession, if included as a CE parameter.

\subsubsection{Precession period}

All optimization tests performed until now used a fixed period $P_\rmn{prec,obs}$, the same  used to build our artificial jet components. But what would happen if we had chosen another value for $P_\rmn{prec,obs}$? 

To tackle this question, we made four optimization runs for T2 data with precession periods corresponding to $0.5$, $0.7$,$1.4$, and $2.0$ of the original value. The CE adopted parameters were   $N=50$, $N_\rmn{elite}=5$, $\alpha=0.6$ and $q=5$. The optimization results for the four periods are presented in Table 4.

We can see that  the minimum value of $S(k_\rmn{max})$ is obtained for the true period; therefore, in a real situation in which this parameter is not well-constrained, we can run the algorithm for different values of the precession period, choosing the solutions that minimises $S(k_\rmn{max})$. Note that the inclusion of the precession period as an additional parameter to be optimised by the algorithm lengthens the completion of the calculations. Our guess is that it might increase the duration of each run by at least a factor of 10. This very rough estimate is based on a set of CE parameters as adopted in this work.

   \begin{figure}
	  {\includegraphics[width = 85 mm, height = 50 mm]{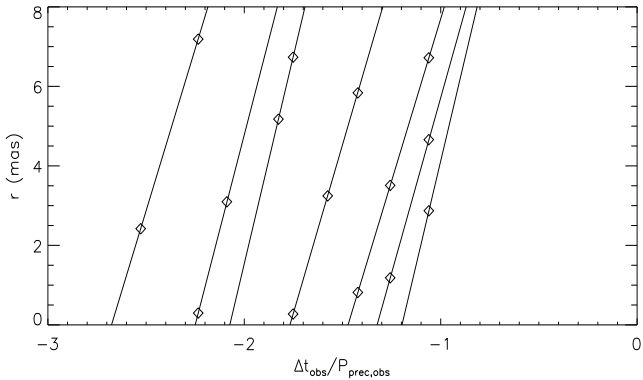}}
      \caption{Core-component distances for test T1. Diamonds represent the synthetic jet components, while lines are the predicted proper motions for each component found from CE precession model optimization. The ejection epochs correspond exactly to those shown in Fig. 2}
      \label{r T1}
   \end{figure}

   \begin{figure}
	  {\includegraphics[width = 85 mm, height = 75 mm]{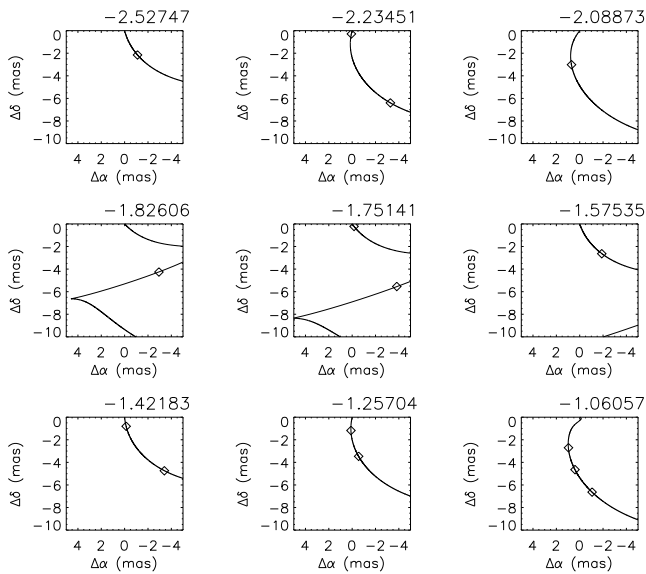}}
      \caption{Snapshots of right ascension and declination offsets of the jet components in test T1 (observation epochs are shown on the top of the plots). Diamonds represent the synthetic jet components, while lines are the predicted instantaneous jet orientation provided by the CE precession model optimization.}
      \label{alpha delta T1}
   \end{figure}

\subsection{Kinematics of the jet components}

In Figs. 10-19 we show the model predictions for the core-component distances, and the right ascension and declination offsets for the T1-T5 tests. Solid lines in these graphs represent the predictions from the CE optimisation adopting $N=100$, $N_\rmn{elite}=5$, $\alpha=0.6$ and $q=5$. Concerning core-component distance plots, the zero-separation epochs refer to those marked by the squares in Figs. 2-4, while the inclinations of the lines correspond to the proper motions obtained from the optimisation processes. In the case of right ascension-declination plots, the continuous lines represent the snapshot of the precession helices generated from the derived precession model parameters.

   \begin{figure}
	  {\includegraphics[width = 85 mm, height = 50 mm]{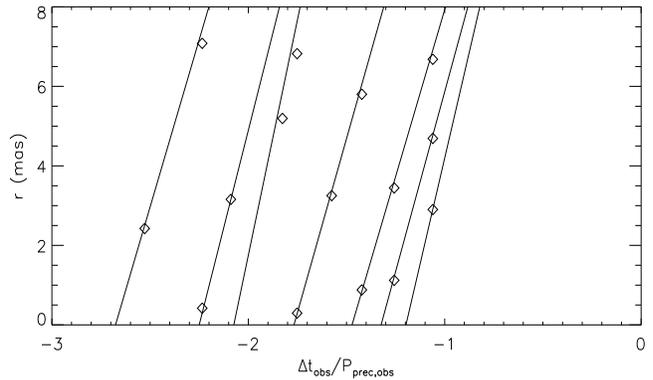}}
      \caption{The same as Fig. 10 but for the precession test T2.}
      \label{r T2}
   \end{figure}

   \begin{figure}
	  {\includegraphics[width = 85 mm, height = 75 mm]{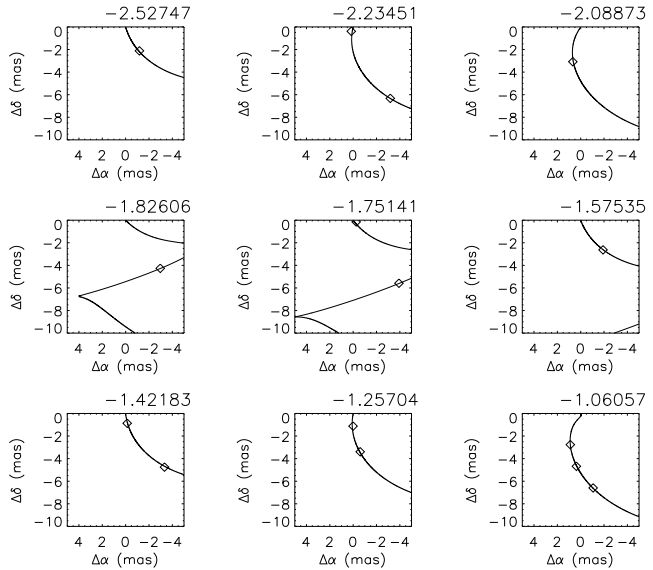}}
      \caption{The same as Fig. 11 but for the precession test T2.}
      \label{alpha delta T2}
   \end{figure}

Note that these calculations did not use any predefined identification scheme for the jet knots, as done in previous works (e.g., \citealt{abca98,caab04a,caab04b}). In other words, it was not provided to the algorithm any information about which pairs of right ascension and declination offsets belong to a given component. It means that misidentification of the jet components due to the lack of a good time sampling of the observations, which might lead to the wrong determination of their kinematic parameters, is completely avoided in our method. Therefore, it relaxes moderately the necessity of collecting data during a long period of time, with the interval between consecutive observations as short as possible.

A good example of how misleading may be the identification of the jet components only based on kinematic features is test T4 presented in Fig. 6. Due to the sparse time coverage, we are induced (specially from the core-component distance plot!) to assume the existence of 7 to 10 non-ballistic jet knots, depending on the data association we made. Obviously their kinetic parameters would differ completely from those calculated considering the correct number of (ballistic) components. It is important to emphasise that we are not claiming the nonexistence of non-ballistic jet motions which have been reported in the literature (e.g., \citealt{hom01,jor05,agu07}). However, some fraction of the bent motions might be the result of a misinterpretation of the observational data due mainly to their bad sampling in time, as the test T4 suggests.

   \begin{figure}
	  {\includegraphics[width = 85 mm, height = 50 mm]{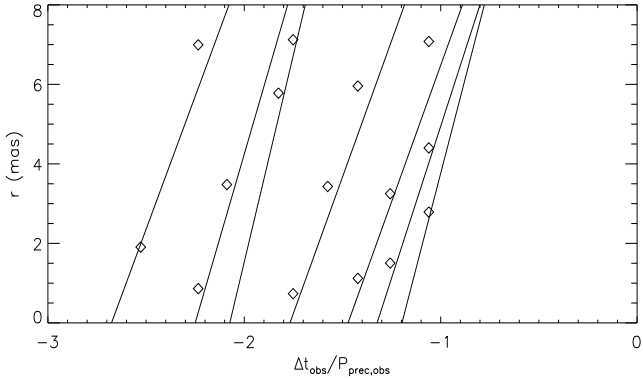}}
      \caption{The same as Fig. 10 but for the precession test T3.}
      \label{r T3}
   \end{figure}

   \begin{figure}
	  {\includegraphics[width = 85 mm, height = 75 mm]{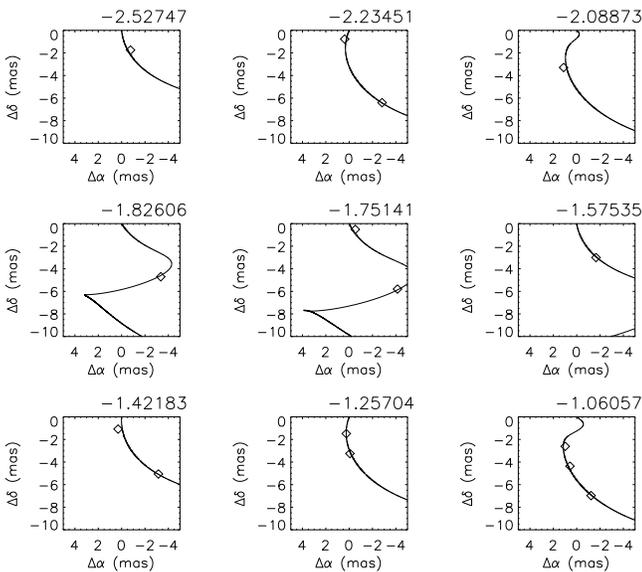}}
      \caption{The same as Fig. 11 but for the precession test T3.}
      \label{alpha delta T3}
   \end{figure}

\section{Conclusions}

Many astrophysical jets present signatures of precession, which were inferred from jet kinematics and/or continuum and line-profile periodic variability (e.g., \citealt{gia73,leva96,caab04a,caab04b}). In the case of jet kinematics, it is necessary to determine previously the apparent proper motions of the knots, as well as their trajectories on the plane of the sky that, on the other hand, depend on the correct identification of the jet components. 

To deal with those potential difficulties, we have implemented the cross-entropy method for continuous multi-extremal optimization with dynamic smoothing \citep{kro06} into our jet precession model (e.g., \citealt{caab04a}). As far as we know, it is the first case of application of the CE technique in astrophysics. 

   \begin{figure}
	  {\includegraphics[width = 85 mm, height = 50 mm]{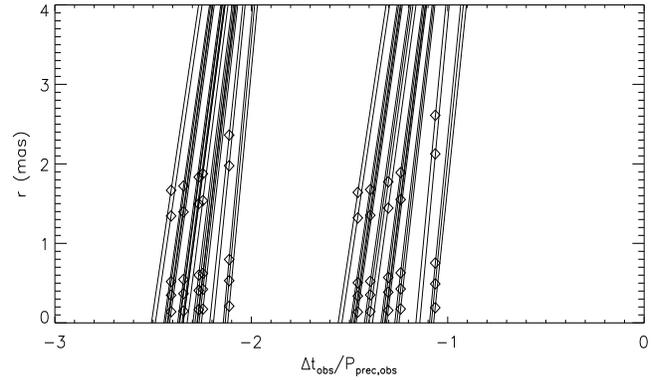}}
      \caption{Core-component distances for test T4. Diamonds represent the synthetic jet components, while lines are the predicted proper motions for each component found from CE precession model optimization. The ejection epochs correspond exactly to those shown in Fig. 3}
      \label{r T4l}
   \end{figure}

   \begin{figure}
	  {\includegraphics[width = 85 mm, height = 105 mm]{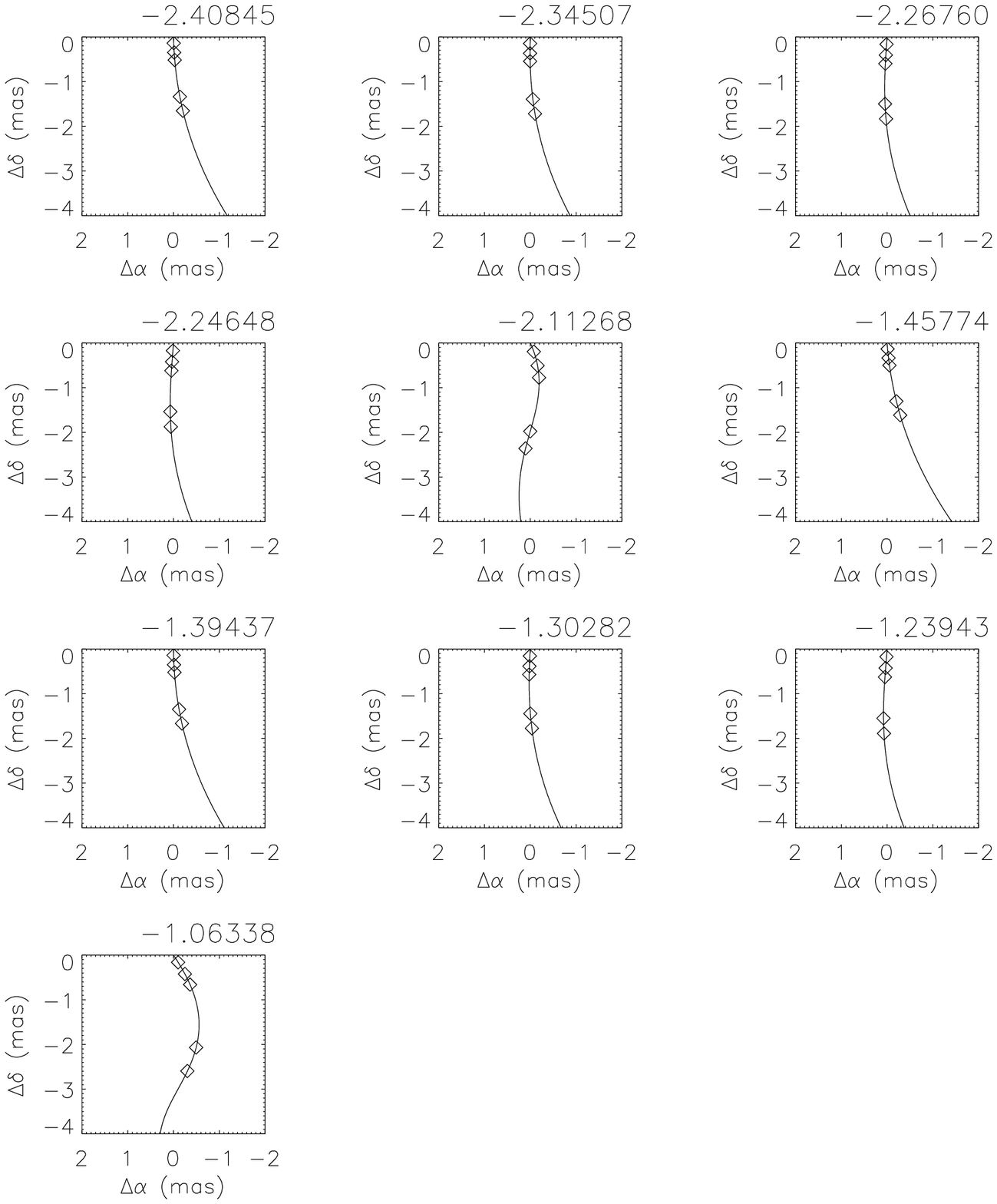}}
      \caption{The same as Fig. 11 but for the precession test T4.}
      \label{alpha delta T4}
   \end{figure}

   \begin{figure}
	  {\includegraphics[width = 85 mm, height = 50 mm]{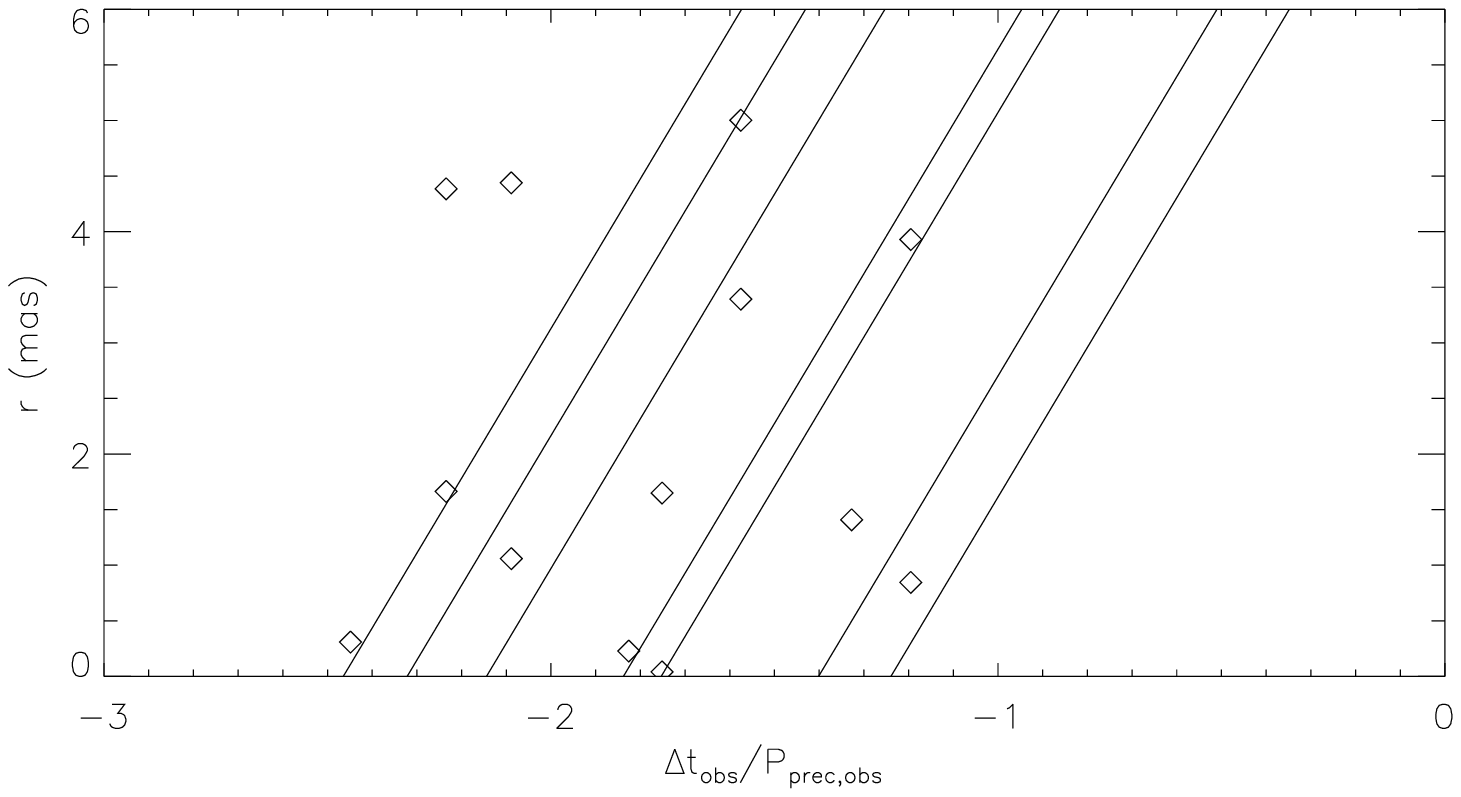}}
      \caption{Core-component distances for test T5. Diamonds represent the synthetic jet components, while lines are the predicted proper motions for each component found from CE precession model optimization. The ejection epochs correspond exactly to those shown in Fig. 4}
      \label{r T5}
   \end{figure}

   \begin{figure}
	  {\includegraphics[width = 85 mm, height = 75 mm]{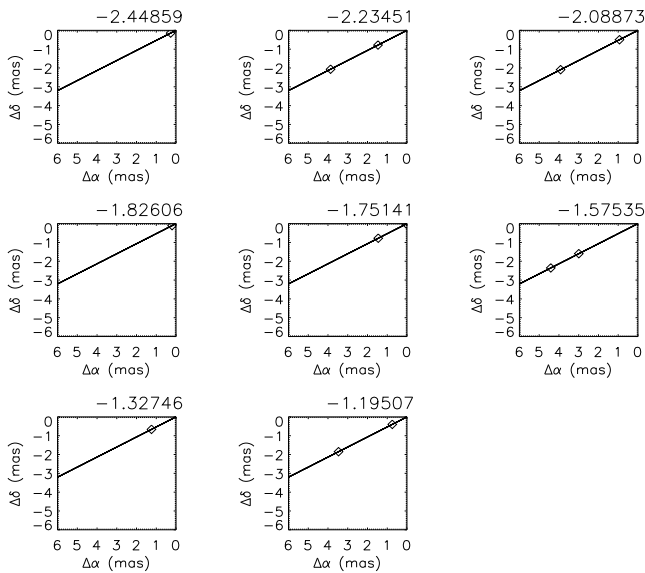}}
      \caption{The same as Fig. 11 but for the precession test T5.}
      \label{alpha delta T5}
   \end{figure}

In this method, tentative sets of precession model parameters are generated randomly from a normal distribution in each iteration, being the best candidates (elite sample) used to construct the next set of solutions. The elite sample was selected from the $N_\rmn{elite}$-set of parameters that better minimises an objective function $S$.

In order to validate our CE precession model, we constructed five sets of data, simulating right ascension and declination offsets between core and jet components, $\Delta\alpha_\rmn{mod}$ and $\Delta\delta_\rmn{mod}$ respectively. These artificial jet components were generated from different combinations of precession model parameters,  assuming that they recede ballistically from the core after their ejection. For each of these tests, the CE precession method was applied to the data, aiming to recover the precession parameters used to create them.

Our results have shown that even in the most challenging tests, the CE method was able to find the correct parameters within $1\%$-level in most cases. The models included highly relativistic precessing jets, as well as a mildly relativistic jet with no precession. The ability to recover the correct precession model parameters was dependent on the cross-entropy parameters used in the optimization: all validation tests performed in this work seems to indicate that $N\ga 50$, $\alpha\ga 0.7$ and $q\approx 5$ improve the algorithm's efficiency. It is important to emphasise that such mapping is necessary since there is no theoretical proposition that points surely to which CE parameters must be used in the calculations to maximise the optimisation's efficiency (e.g., \citealt{kro06}).

Component identification problems related to fitting procedures, as well as observations sparsely sampled in time and sources with high rate production of jet components, may influence  the follow up of the jet knots, which consequently might contribute to a kinematic misinterpretation of the data. In order to verify how sensitive is our algorithm to this issue, we performed the T4-test that consists of 50 precessing jet components generated to mimic eight non-ballistic knots. Our technique was able to recover the original jet precession parameters, ignoring any previous components misidentification. Note that the inclusion of random fluctuations in the right ascension and declination offsets decreases the ability of the algorithm in recovering the real precession parameters.

To keep the numerical execution of the algorithm within acceptable intervals, we fixed the precession period in the observer's reference frame and the sense of precession (clockwise or counterclockwise rotation) during the optimization processes. This practical limitation led us to perform additional tests to check the behaviour of the CE method when a wrong choice for the precession period or the sense of rotation is made. As expected, such additional runs clearly indicated that the minimum value of the objective function is obtained when the correct values for the precession period and sense of rotation are employed.

Our method was successful in pointing out the lack of precession in the case of test T5 (mildly relativistic non-precessing jet), recovering the values of $\varphi_0$ and $\eta_0$ with a very impressive precision. Although reasonable values have been also obtained for $\gamma$ and $\Delta\tau_\rmn{s}$, the method failed completely to find the correct value for $\phi_0$. This is not supposed to be a serious issue since we are modelling a non-precessing jet in terms of a precession model, which obviously can produce some inconsistencies in the results. Nevertheless, we argue that our CE method can be useful in indicating the presence of any precession signature in the observational data.

Our results show the great potentiality of the CE method to study precession (or the lack of) in astrophysical jets, utilising only the time variation of the right ascension and declination coordinates of the knots.

\section*{Acknowledgments}

This work was supported by the Brazilian Agencies FAPESP and CNPq. The authors also thank the anonymous referee for careful reading of the manuscript and for useful comments and suggestions.

\bsp

\label{lastpage}

\end{document}